\documentclass[11pt]{article} 
\usepackage{graphicx}
\newcommand{\be}{\begin{equation}}
\newcommand{\ee}{\end{equation}}
\newcommand{\bea}{\begin{eqnarray}}
\newcommand{\eea}{\end{eqnarray}}

\begin{document}
\vspace{0.5in}
\begin{center}
{\LARGE{\bf Wide Class of Logarithmic Potentials with Power-Tower Kink Tails}}
\end{center}

\vspace{0.2in} 

\begin{center}
{{\bf Avinash Khare}} \\
{Physics Department, Savitribai Phule Pune University \\
 Pune 411007, India}
\end{center}

\begin{center}
{{\bf Avadh Saxena}} \\ 
{Theoretical Division and Center for Nonlinear Studies, 
Los Alamos National Laboratory, Los Alamos, New Mexico 87545, USA}
\end{center}

\vspace{0.2in}

\noindent{\bf {Abstract:}}

We present a wide class of potentials which admit kinks and corresponding 
mirror kinks with either a power law or an exponential tail at the two 
extreme ends 
and a power-tower form of tails at the two neighbouring ends, i.e. of the forms
$ette$ or $pttp$ where $e, p$ and $t$ denote exponential, power law and 
power-tower tail, respectively. We analyze kink stability equation in all 
these cases and show that there is no gap between the zero mode and the 
beginning of the continuum. Finally, we provide a recipe for obtaining logarithmic 
potentials with power-tower kink tails and estimate kink-kink interaction strength. 

\section{Introduction} 

A vast majority of kink solutions obtained during the last four decades for 
a variety of field theory potentials, e.g. sine-Gordon, 
double sine-Gordon, $\phi^4$, $\phi^6$, etc. harbor kinks with an 
exponential tail \cite{Raj}. Recently we and others have presented a wide class of 
kink-bearing potentials for which one has a power law kink tail \cite{KCS, Chapter, KS, gomes, bazeia}. 
Very recently we have also presented a model with a super-exponential profile 
with one of the tails also being 
super-exponential 
\cite{KKS, menezes, marques, bazeia2}.  Thus by now we have models where one
has a variety of kink tails such as of power law, exponential or super-exponential form.  
The obvious question is if there are models with still different types of kink
tails. The purpose of this paper is to present an entirely different and 
novel class of potentials with {\it power-tower} kink tails, thus further expanding 
the type of kink asymptotes one could realize. One of the logarithmic potentials with 
super-exponential kink 
tails arises in the context of infinite order phase transitions \cite{KKS} and,  
therefore, conceivably the family of potentials considered here may have similar 
physical relevance.

The plan of the paper is as follows.  In Section 2 we discuss a one-parameter 
family of potentials with kink tails 
of the form $ette$.  We discuss the stability analysis of these kink solutions, 
which are expressed in terms of the exponential integral function $Ei(x)$ 
\cite{erdelyi, grad} and show that there is no gap between the zero mode and the 
beginning of the continuum. In Section 3 
we consider a two-parameter family of potentials which lead to kink tails of the 
form $pttp$.  We also discuss the stability analysis of these solutions and
show that even in this case there is no gap between the zero mode and the 
beginning of the continuum. Our main conclusions are summarized in Section 4 
where we also discuss how to obtain potentials with power-tower kink tails and 
estimate interaction between such kinks.

\section{Models With Tails of the Form $ette$}

In this section we consider a one parameter family of potentials of the form
\be\label{2.1}
V(\phi) = (1/2) \phi^{2m+2} [(1/2) \ln(\phi^2)]^2\,,~~m \ge 1\,.
\ee 
These potentials have degenerate minima at $\phi = 0, \pm 1$ with 
$V_{min} = 0$ while they have degenerate maxima at
\be\label{2.1a}
\phi_{max} = \pm \, e^{-1/(m+1)}\,,~~V_{max} = \frac{1}{2e^2 (m+1)^2}\,.
\ee
Thus notice that while $\phi_{max}(m = 1) = \pm e^{-1/2}$, as $m$ becomes
larger, then $\phi_{max}$ moves towards $\pm 1$. On the other hand 
while, $V_{max}(m =1) = \frac{1}{8e^2}$, as $m$ becomes larger, $V_{max}$ 
decreases progressively towards zero. All these models for any integer $m$  
admit a kink from $0$ to $1$ and a mirror kink from $-1$ to
$0$ (and corresponding antikinks) with
tails of the form $ette$. As an illustration we first discuss the case of 
$m = 1, 2$ and then generalize to arbitrary $m$. The potential given by Eq. (1) is shown 
in Fig. 1 for different values of $m$. 

\begin{figure}[h] 
\includegraphics[width= 5.1 in]{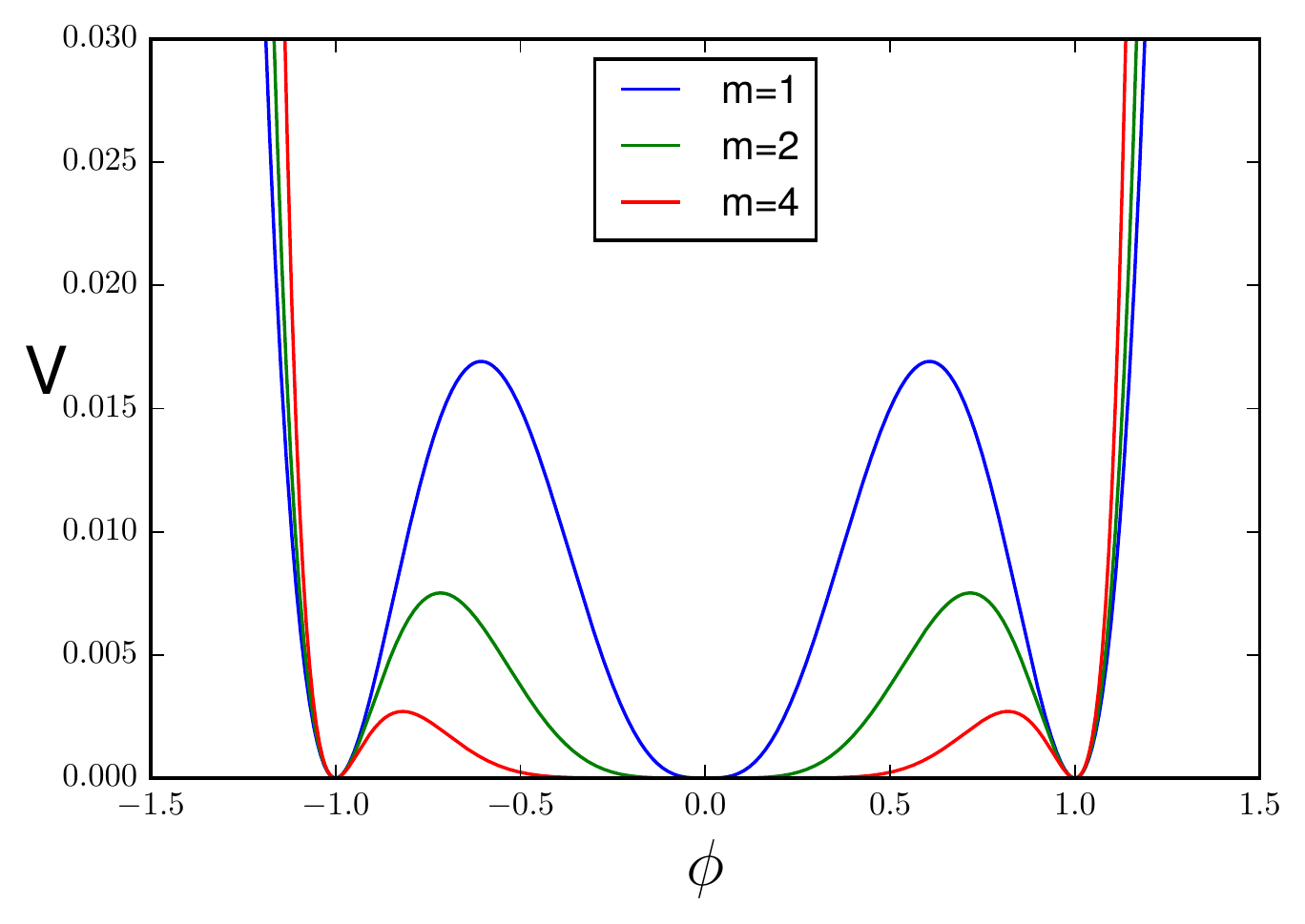}
\caption{Potential $V(\phi)$ for $m=1$, $m=2$ and $m=4$ (see Eq. (1)).  }
\end{figure}  

\vspace{0.1in} 
\noindent{\bf Case I: $m = 1$}

Consider the potential
\be\label{2.2}
V(\phi) = (1/2) \phi^{4} [(1/2) \ln(\phi^2)]^2\,.
\ee 
Thus the self-dual equation we need to solve is 
\be\label{2.3}
\frac{d\phi}{dx} = \pm \phi^2 [(1/2) \ln(\phi^2)]\,.
\ee
For the kink between $0$ and $1$ we need to solve the self-dual Eq. (\ref{2.3})
with negative sign. In fact this is true no matter what $m$ ($\ge 1$) is. 
This is easily integrated by making the substitution $t = (1/2)\ln(\phi^2)$
and we obtain the implicit kink solution
\be\label{2.4}
-x = \int \frac{e^{-t}}{t}\, dt = Ei(-t)\,,
\ee
where $Ei(x)$ denotes the exponential integral function \cite{erdelyi, grad}. 
Unfortunately, we do not know how to invert this function analytically \cite{inverse} and 
obtain $t$ and hence $\phi$ in terms of $x$. However, using the Taylor series 
expansion of $Ei(x)$ as given in \cite{erdelyi} 
\be\label{2.5}
Ei(x) = \gamma +\ln|x| + x + \frac{x^2}{2 2!} +...\,,
\ee
as well as the asymptotic formula \cite{erdelyi}
\be\label{2.6}
Ei(x) \equiv e^{x} \left[\frac{1}{x} +\frac{1}{x^2} +\frac{2!}{x^3} 
+\frac{3!}{x^4}+...\right]\,,
\ee
we can estimate the tail behaviour around $\phi = 0$ as $x \rightarrow -\infty$
and around $\phi = 1$ as $x \rightarrow +\infty$. Here $\gamma$ = 0.577 is 
the Euler's constant.

\begin{figure}[h] 
\includegraphics[width= 5.1 in]{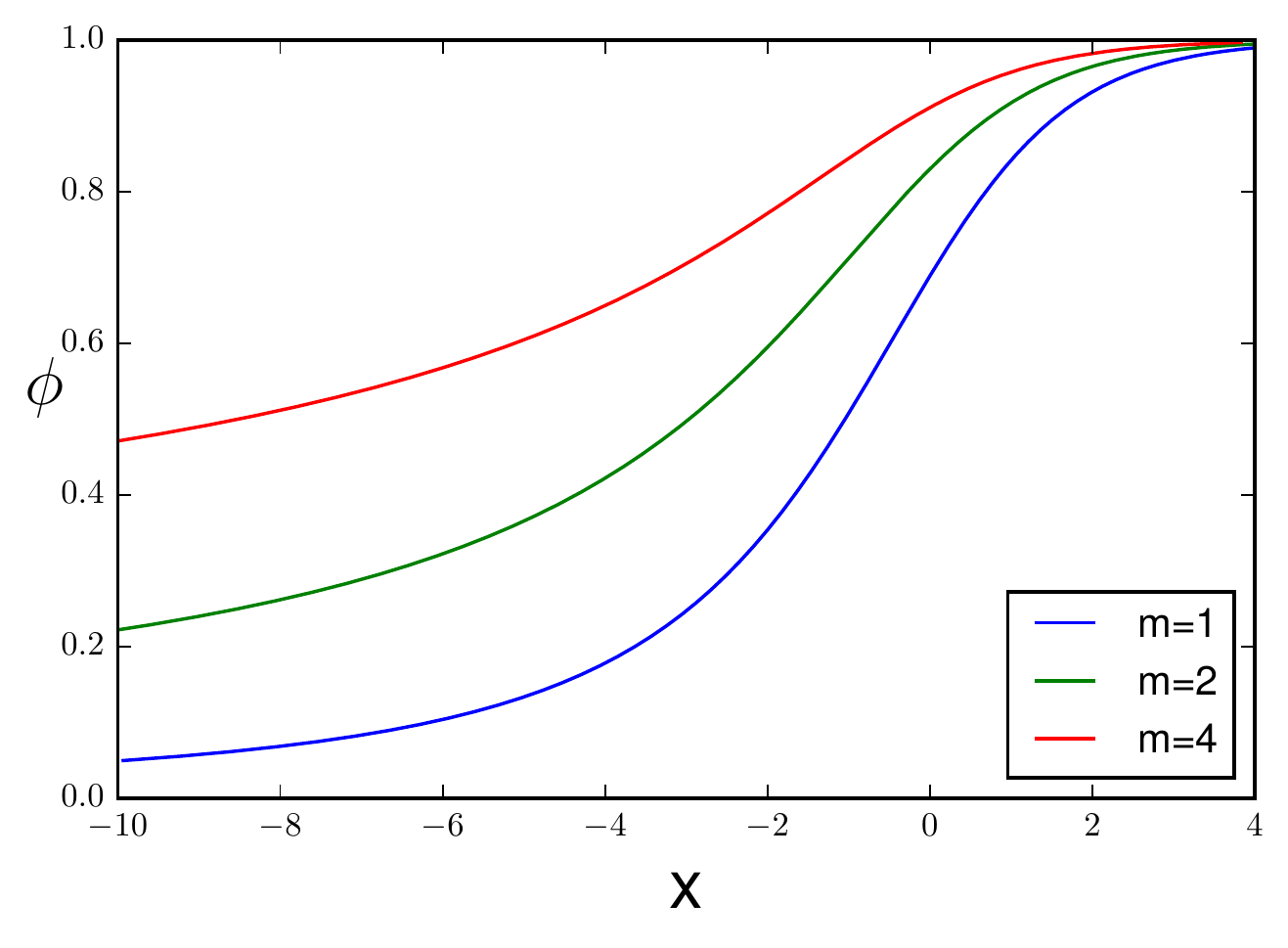}
\caption{Kink solution $\phi(x)$ for $m=1$, $m=2$ and $m=4$ (see Eq. (17)).  }
\end{figure}  

We find that
\be\label{2.7}
\lim_{x \rightarrow -\infty} \phi(x) \ln[\phi(x)] = \frac{1}{x}\,,~~
\lim_{x \rightarrow +\infty} \phi(x) = 1- e^{-(x+\gamma)}\,.
\ee 
It is worth pointing out that the asymptotic behaviour around $\phi = 0$
(as $x \rightarrow -\infty$) can also be written as 
\be\label{2.8}
\lim_{x \rightarrow -\infty} \phi(x)^{\phi(x)} = e^{1/x}\,,
\ee
which is known in the literature as the power-tower function \cite{powert} of order two.  
It is also related to the iterated or repeated exponentiation, i.e.~tetration \cite{tetration}.  
We can invert Eq. (5) numerically and obtain the kink solution as 
given in Fig. 2.  For $x\rightarrow\infty$ the kink tail approaches $\phi=1$ 
asymptotically as an {\it exponential} tail whereas for  $x\rightarrow-\infty$ 
it approaches $\phi=0$ asymptotically as a {\it power-tower} tail. 

\vspace{0.1in}  
\noindent{\bf Case II: $m = 2$}

Consider the potential
\be\label{2.8a}
V(\phi) = (1/2) \phi^{6} [(1/2) \ln(\phi^2)]^2\,.
\ee 
Thus the self-dual equation we need to solve is 
\be\label{2.9}
\frac{d\phi}{dx} = \pm \phi^3 [(1/2) \ln(\phi^2)]\,.
\ee
This is easily integrated by making the substitution $t = (1/2)\ln(\phi^2)$
and we obtain the implicit kink solution
\be\label{2.10}
-x = \int \frac{e^{-2t}}{t}\, dt = Ei(-2t)\,.
\ee
From here using Eqs. (\ref{2.5}) and (\ref{2.6}) it is easily checked that
\be\label{2.11}
\lim_{x \rightarrow -\infty} \phi^2(x) \ln[\phi(x)] = \frac{1}{2x}\,,~~
\lim_{x \rightarrow +\infty} \phi(x) = 1- \frac{1}{2} e^{-(x+\gamma)}\,.
\ee 
It is worth pointing out that the asymptotic behaviour around $\phi = 0$
(as $x \rightarrow -\infty$) in Eq. (\ref{2.11}) can also be written as 
\be\label{2.12}
\lim_{x \rightarrow -\infty} \phi(x)^{{\phi(x)}^{2}} = e^{1/2x}\,,
\ee
which is known in the literature as the power-tower function of order two \cite{powert} 
or tetration \cite{tetration}.  The kink solution is obtained by numerically inverting Eq. (12) as 
shown in Fig. 2. 

\vspace{0.1in}  
\noindent{\bf Case III: Arbitrary $m$}

The generalization to arbitrary $m$ is now straightforward.
Consider the potential
\be\label{2.13}
V(\phi) = (1/2) \phi^{2m+2} [(1/2) \ln(\phi^2)]^2\,.
\ee 
Thus the self-dual equation we need to solve is 
\be\label{2.14}
\frac{d\phi}{dx} = \pm \phi^{m+1} [(1/2) \ln(\phi^2)]\,.
\ee
This is easily integrated by making the substitution $t = (1/2)\ln(\phi^2)$
and we obtain the implicit kink solution
\be\label{2.15}
-x = \int \frac{e^{-mt}}{t}\, dt = Ei(-mt)\,.
\ee
From here using Eqs. (\ref{2.5}) and (\ref{2.6}) it is easy to see that
\be\label{2.16}
\lim_{x \rightarrow -\infty} \phi^{m}(x) \ln[\phi(x)] = \frac{1}{mx}\,,~~
\lim_{x \rightarrow +\infty} \phi(x) = 1- \frac{1}{m} e^{-(x+\gamma)}\,.
\ee 
It is worth pointing out that the asymptotic behaviour around $\phi = 0$
(as $x \rightarrow -\infty$) in Eq. (\ref{2.16}) can also be written as 
\be\label{2.17}
\lim_{x \rightarrow -\infty} \phi(x)^{{\phi(x)}^{m}} = e^{1/mx}\,,
\ee
which is known in the literature as the power-tower function of order two \cite{powert} 
or tetration \cite{tetration}.  Kink profiles for three different values of $m$ are depicted 
in Fig. 2.  Note that with increasing $m$, the approach to $\phi=0$ for large 
negative $x$ becomes 
progressively slower in accordance with the power-tower function.  In other words, for 
large $m$ the kink profile tends to become symmetric. 

\vspace{0.1in}  
\noindent{\bf Kink Mass:}

One can easily calculate the kink mass for the entire family of potentials. In particular,
for the kink potential as given by Eq. (\ref{2.13}), the kink mass is given by
\be\label{2.16a}
M_K = \int_{0}^{1} \sqrt{2V(\phi)}\, d\phi = \int_{0}^{1} d\phi\, 
\phi^{m+1} \ln(\phi) d\phi = \frac{1}{(m+2)^2}\,.
\ee
Note that the kink mass decreases as $m$ increases.

\subsection{Stability Analysis}

We can perform the stability analysis of all the above kink solutions and show that, 
akin to the kinks with the power law tail \cite{KS}, for all the above kink solutions, 
there is no gap between the zero mode and the beginning of the continuum.

As an illustration, we discuss the $m = 1$ case in detail, the generalization
to the arbitrary $m$ case is then straightforward.
In this case the self-dual equation is as given by Eq. (\ref{2.3}) with minus sign.
Thus the kink zero mode is given by
\be\label{2.20}
\psi_0(x) = \frac{d\phi_k}{dx} \propto \left[(\phi_k(x)\right]^2 \ln[\phi_k(x)]\,,
\ee
where $\phi_k$ is the kink solution. The above zero mode $\psi_0$ is  clearly 
nodeless and vanishes
as $x \rightarrow \pm \infty$ since as $x$ goes from $-\infty$ to $\infty$,
$\phi$ varies from $0$ to $1$. 

We can also calculate the corresponding kink potential $V_K(x)$ which appears in the
stability equation
\be\label{2.21}
-\frac{d^2\psi}{dx^2} + V_K(x) \psi = \omega^2 \psi\,,
\ee
where $V_K(x) = \frac{d^2V(\phi)}{d\phi^2}$, evaluated at $\phi = \phi_k(x)$.
On using the potential for $m = 1$ as given by Eq. (\ref{2.2}) we find
that
\be\label{2.22}
V_K(x) = \frac{d^2V(\phi_k)}{d\phi^2} = 6(\phi_k)^2 [\ln(\phi_k)]^2 
+ 7 (\phi_k)^2 \ln(\phi_k) + (\phi_k)^2\,, 
\ee
and hence it is clear that while $V_K(x = \infty) = 1$, $V_K(-\infty) = 0$.
Thus the continuum begins at $\omega^2 = 0$, i.e. there is no gap between
the zero mode and the beginning of the continuum.

The generalization to the arbitrary $m$ case is now straightforward. In
particular, we find that the kink zero mode is given by
\be\label{2.23}
\psi_0(x) = \frac{d\phi_k}{dx} = [\phi_k(x)]^{m+1} \ln[\phi_k(x)]\,,
\ee
which is clearly nodeless. The corresponding kink stability potential
is given by
\bea\label{2.24}
&&V_K(x) = \frac{d^2V(\phi_k)}{d\phi^2} = (m+1)(2m+1)(\phi_k)^{2m} 
[\ln(\phi_k)]^2 \nonumber \\ 
&&+ (4m+3) (\phi_k)^{2m} \ln(\phi_k) + (\phi_k)^{2m}\,, 
\eea
from where again it is clear that there is no gap between the zero mode and the
beginning of the continuum. 

\subsection{Nature of Kink-Kink and Kink-Antikink Interactions}

In this section we have obtained kink  and mirror kink solutions (and 
corresponding antikinks) for a one-parameter family
of potentials as given by Eq. (\ref{2.1}). The kinks are from $0$ to $1$ (and
mirror kinks are from $-1$ to $0$) as $x$ goes from $-\infty$ to $\infty$.
We have seen that while the kink tail around $\phi = 1$ or $\phi = -1$ is
exponential, the kink tail around $\phi = 0$ has a power-tower form.  
Using this information, let us try to qualitatively understand the nature of
kink-kink (KK) and kink-antikink (K-AK) interactions. 

Let us first consider the kink-kink interaction between the $(-1,0)$ mirror 
kink and the $(0,1)$ kink. From Eq. (\ref{2.1}) it is clear that around 
$\phi = 0$ the kink potential is as given by Eq. (\ref{2.1}). Notice that
if there were no $\ln(\phi^2)$ term in Eq. (\ref{2.1}) then using the recent
approach of Manton for potentials with a power law tail \cite{man, ivan} one
would have immediately predicted that the KK force would be  
\be\label{2.25}
F_{KK} = \bigg [\frac{\Gamma[m/2(m+1)] \Gamma[1/2(m+1)]}
{2m \sqrt{\pi}} \bigg ]^{2(m+1)/m} \frac{1}{2 R^{2(m+1)/m}}\,,
\ee 
where $R$ is the distance  between the two kinks. So the question is: what is 
the effect of the $[(1/2)\ln(\phi^2)]$ term multiplying the potential in Eq. (\ref{2.1})? 
In this connection we notice that in a recent publication by
the present authors \cite{KKS} we have shown that in the case of the potential
$V(\phi) = (1/2) \phi^2 [(1/2)\ln(\phi^2)]^2$, while the KK 
force would have been exponentially small if there were no $\ln(\phi^2)$ term 
present, because of the $\ln(\phi^2)$ term, the force actually gets even weaker
and is in fact super-exponential. Taking this as a guide, we would 
therefore expect that the KK force in the case of potential (\ref{2.1}) will 
still have a power law fall off but perhaps with a slower fall off and the 
strength of 
the force too would be different. Only either a new 
formalism or numerical estimation can decide the issue. 

Similar conclusion is also true concerning the force between $(1,0)$ AK and
$(0,1)$ K. On the other hand the force between $(0,1)$ K and $(1,0)$ 
AK will be exponentially small and using the original Manton formalism 
\cite{manton} is of the form $e^{-R}$.

\section{Models With Tails of the Form $pttp$}

In this section we present a two-parameter family of potentials of the form
\be\label{3.1}
V(\phi) = (1/2) \phi^{2m+2} [(1/2) \ln(\phi^2)]^{2n+2}\,,~~m, n \ge 1\,.
\ee 
These potentials have degenerate minima at $\phi = 0, \pm 1$ with 
$V_{min} = 0$ while they have degenerate maxima at
\be\label{3.1a}
\phi_{max} = \pm \, e^{-(n+1)/(m+1)}\,,~~
V_{max} = \frac{1}{2e^{2(n+2)}} [\frac{(n+1)}{(m+1)}]^{2(n+1)}\,.
\ee
Notice that both $\phi_{max}$ and $V_{max}$ depend on two parameters $m$ and
$n$. Notice also that for a fixed $m$, as $n \rightarrow \infty$, $\phi_{max}
\rightarrow 0$ and $V_{max} \rightarrow 0$. On the other hand, for a fixed
$n$, as $m \rightarrow \infty$, $\phi_{max} \rightarrow 1$ and 
$V_{max} \rightarrow 0$. Finally, for $m = n$, $\phi_{max} = \pm e^{-1}$
and the corresponding $V_{max} = \frac{1}{2e^{2(n+1)}}$.  
It is interesting to note that for a given $m$, all the potentials as 
given by Eq. (\ref{3.1}) with arbitrary integral $n$ have the same value
$V(\phi) = \frac{1}{2e^{2(m+1)}}$ in case $\phi = \pm 1/e$  or 
$V(\phi) = \frac{e^{2(m+1)}}{2}$ in case $\phi = \pm e$.

All these models for any integers $m$  and $n$  
admit a kink from $0$ to $1$ and a mirror kink from $-1$ to
$0$ (and corresponding antikinks) with tails of the form $pttp$. As an 
illustration we first discuss the case of 
arbitrary $m$ and $n = 1, 2$ and then generalize to arbitrary $n$.

\vspace{0.1in}  
\noindent{\bf Case I: $n = 1, m~Arbitrary$}

Consider the potential
\be\label{3.2}
V(\phi) = (1/2) \phi^{2m+2} [(1/2) \ln(\phi^2)]^4\,.
\ee 
The potential in Eq. (29) is shown in Fig. 3 for different values of $m$. 
Thus the 
self-dual equation we need to solve is 
\be\label{3.3}
\frac{d\phi}{dx} = \pm \phi^{m+1} [(1/2) \ln(\phi^2)]^2\,.
\ee

\begin{figure}[h] 
\includegraphics[width= 5.1 in]{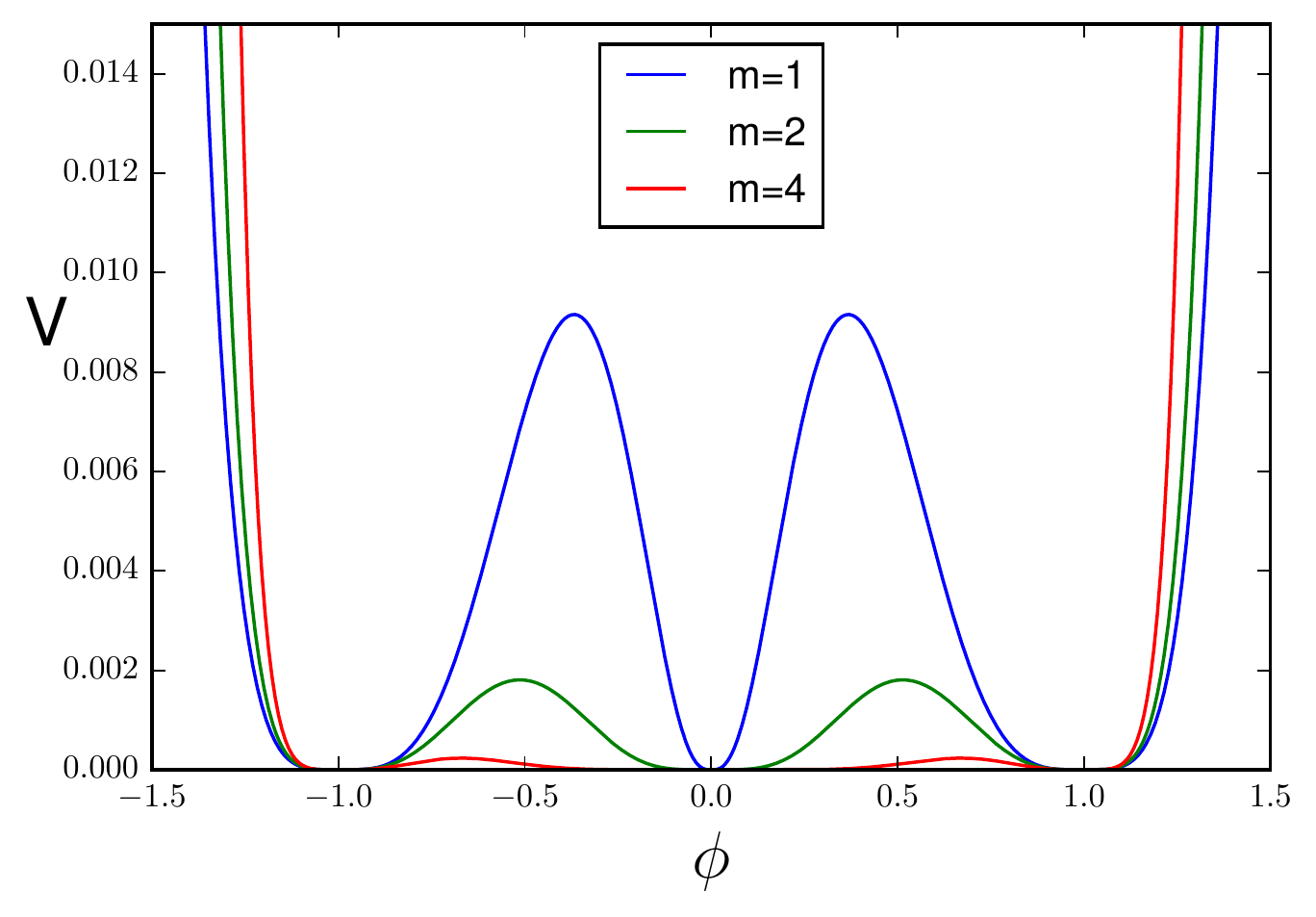}
\caption{Potential $V(\phi)$ for $n=1$ and $m=1$, $m=2$ and $m=4$ (see Eq. (29)).  }
\end{figure}   

For the kink between $0$ and $1$ we need to solve the self-dual Eq. (\ref{3.3})
with positive sign.  
This is easily integrated by making the substitution $t = (1/2)\ln(\phi^2)$
and we obtain the implicit kink solution
\be\label{3.4}
x = \int \frac{e^{-mt}}{t^2}\, dt = -\frac{e^{-mt}}{t} -m\,Ei(-mt)\,.
\ee
Using Eqs. (\ref{2.5}) and (\ref{2.6}) we then find that
\be\label{3.5}
\lim_{x \rightarrow -\infty} \phi(x)^{m} \big(\ln[\phi(x)]\big)^2 
= -\frac{1}{mx}\,,~~
\lim_{x \rightarrow +\infty} \phi(x) = 1- \frac{1}{(x+m\gamma)}\,.
\ee 
It is worth pointing out that the asymptotic behaviour around $\phi = 0$
(as $x \rightarrow -\infty$) can also be written as 
\be\label{3.6}
\lim_{x \rightarrow -\infty} \phi(x)^{{\phi(x)}^{m/2}} 
= e^{1/\sqrt{-mx}}\,.
\ee

\begin{figure}[h] 
\includegraphics[width= 5.1 in]{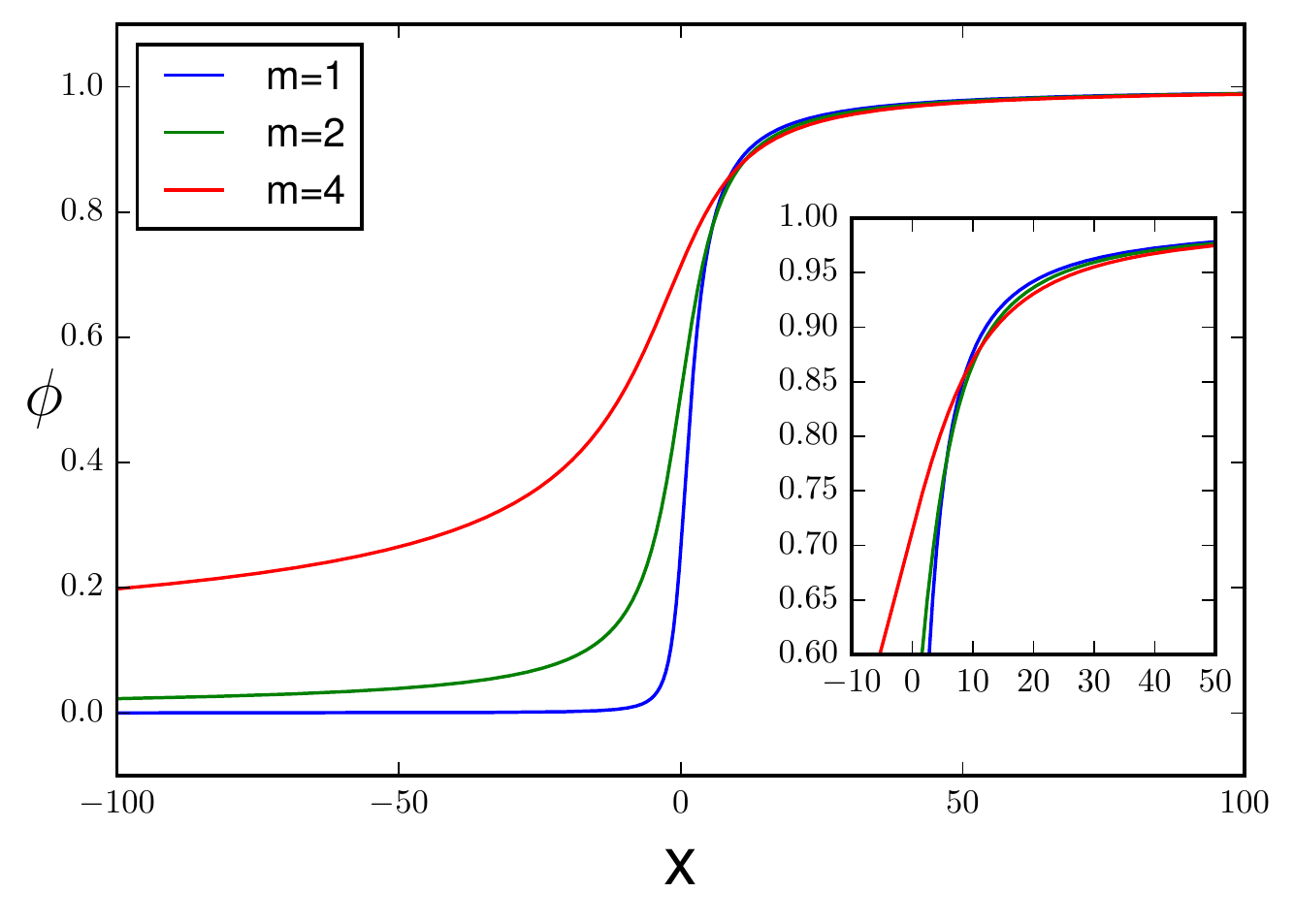}
\caption{Kink solution $\phi(x)$ for $n=1$ and $m=1$, $m=2$ and $m=4$ (see Eq. (31)). Inset 
shows in detail how the three profiles cross each other for $x>0$. }
\end{figure} 

Equation (\ref{3.4}) can be inverted numerically and the kink profile is depicted in Fig. 4. 
For $x\rightarrow\infty$ the kink tails approach $\phi=1$ as a power-law whereas 
for $x\rightarrow-\infty$ the kink tails approach $\phi=0$ as a power-tower function. 
With increasing $m$ for large negative $x$ the tails approach $\phi=0$ 
progressively slowly.  In 
other words, for large $m$ the kink profile tends to become symmetric. 

\vspace{0.1in}  
\noindent{\bf Case II: $n = 2$}

Consider the potential
\be\label{3.7}
V(\phi) = (1/2) \phi^{2m+2} [(1/2) \ln(\phi^2)]^6\,.
\ee 
Thus the self-dual equation we need to solve is 
\be\label{3.8}
\frac{d\phi}{dx} = \pm \phi^{m+1} \big([(1/2) \ln(\phi^2)]\big)^{3}\,.
\ee
For the kink between $0$ and $1$, unlike the $n = 1$ case, we need to 
solve the self-dual Eq. (\ref{3.8})
with negative sign.  
This is easily integrated by making the substitution $t = (1/2)\ln(\phi^2)$
and we obtain the implicit kink solution
\be\label{3.9}
-x = \int \frac{e^{-mt}}{t^3}\, dt 
= -\frac{e^{-mt}}{2t^2} + \frac{m e^{-mt}}{2t} +\frac{m^2}{2} Ei(-mt)\,.
\ee
Using Eqs. (\ref{2.5}) and (\ref{2.6}) we then find that
\be\label{3.10}
\lim_{x \rightarrow -\infty} \phi^{m}(x) \big(\ln[\phi(x)]\big)^{3} 
= \frac{1}{mx}\,,~~
\lim_{x \rightarrow +\infty} \phi(x) = 1- \frac{1}{[2x+m^2\gamma]^{1/2}}\,.
\ee 
It is worth pointing out that the asymptotic behaviour around $\phi = 0$
(as $x \rightarrow -\infty$) in Eq. (\ref{3.10}) can also be written as 
\be\label{3.11}
\lim_{x \rightarrow -\infty} \phi(x)^{{\phi(x)}^{m/3}} = e^{1/(-mx)^{1/3}}\,,
\ee
which is known in the literature as the power-tower function of order two 
\cite{powert} or tetration \cite{tetration}. 

\vspace{0.1in}  
\noindent{\bf Case III: Arbitrary $n$}

The generalization to arbitrary $n$ is now straightforward.
Consider the potential
\be\label{3.12}
V(\phi) = (1/2) \phi^{2m+2} [(1/2) \ln(\phi^2)]^{2n+2}\,, 
\ee 
which is shown in Fig. 5 for $m=1$ and general $n$.  Thus the self-dual 
equation we need to solve is 
\be\label{3.13}
\frac{d\phi}{dx} = \pm \phi^{m+1} \big([(1/2) \ln(\phi^2)]\big)^{n+1}\,.
\ee

\begin{figure}[h] 
\includegraphics[width= 5.1 in]{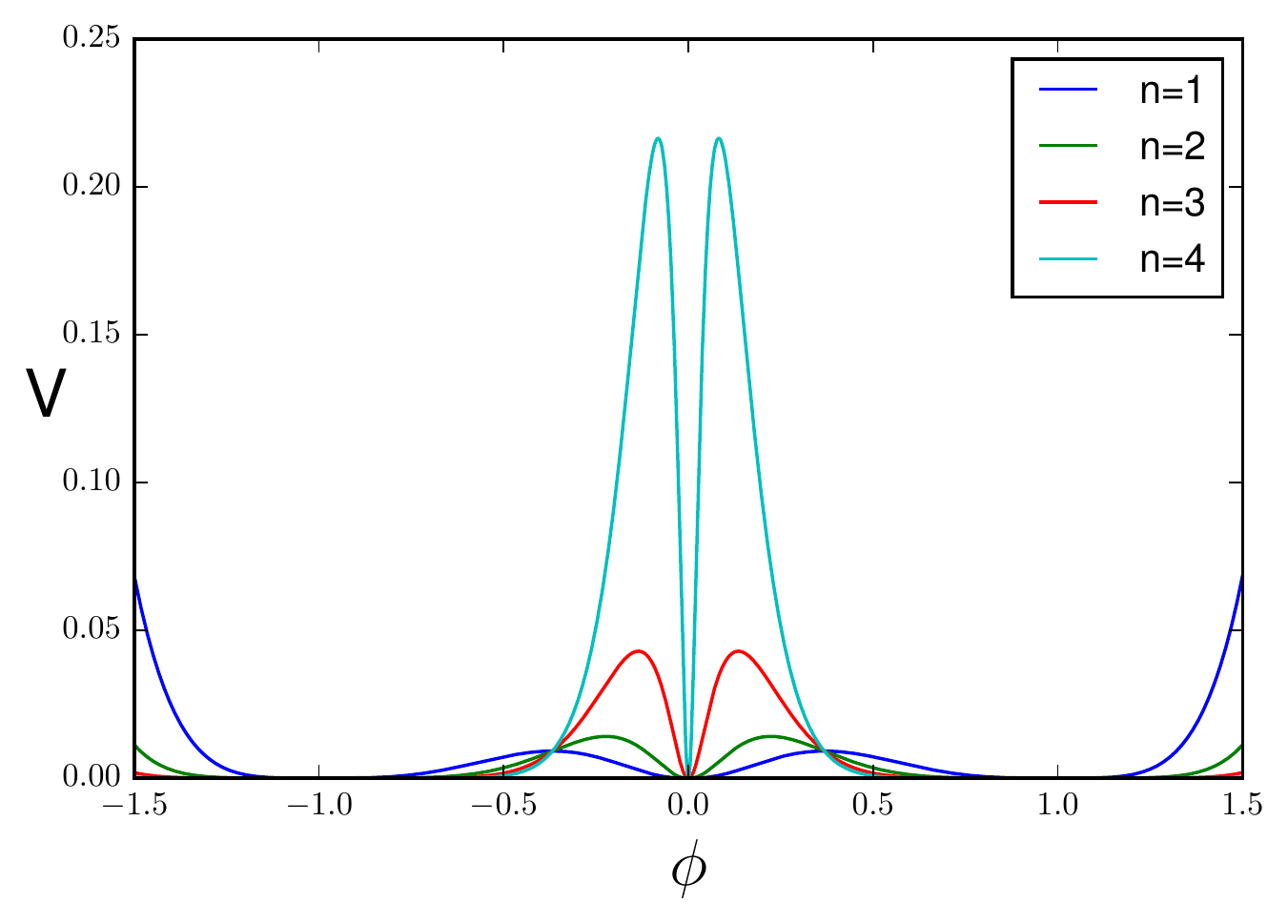}
\caption{Potential $V(\phi)$ for $m=1$, and $n=1$, $n=2$, $n=3$ and $n=4$ (see Eq. (39)).  }
\end{figure} 
This is easily done by making the substitution $t = (1/2)\ln(\phi^2)$
and we obtain 
\be\label{3.14}
\pm x = \int \frac{e^{-mt}}{t^{n+1}}\, dt\,.
\ee
This is easily integrated using \cite{grad} 
\be\label{3.15}
\int \frac{e^{ax}}{x^{n}}\, dx = -e^{ax}  \sum_{k= 1}^{k = n-1} 
\frac{a^{k-1}}{(n-1)(n-2)... (n+1-k) x^{n-k}} +\frac{a^{n-1}}{(n-1)!} Ei(ax)\,.
\ee
We obtain
\be\label{3.16}
\pm x = -e^{-mt}  \sum_{k=1}^{n} \frac{(-m)^{k-1}}{n(n-1)...(n+1-k) 
t^{n+1-k}} +\frac{(-m)^{n}}{n!} Ei(-mt)\,, 
\ee 
which can be numerically inverted to obtain the kink solutions, as depicted in Fig. 6 
for $m=1$ and different values of $n$.   

\begin{figure}[h] 
\includegraphics[width= 5.1 in]{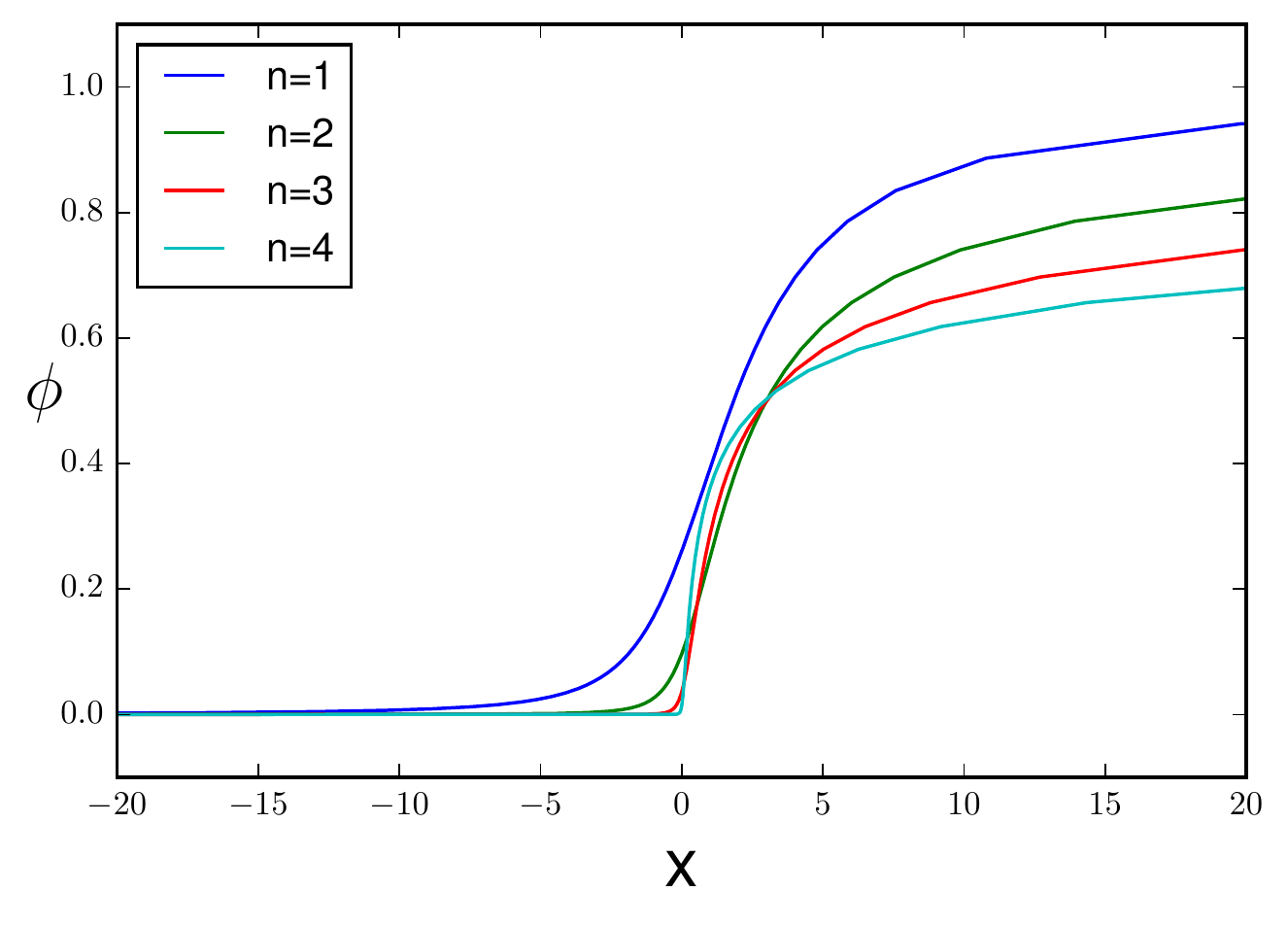}
\caption{Kink solution $\phi(x)$ for $m=1$ and $n=1$, $n=2$, $n=3$ and $n=4$ (see Eq. (43)). }
\end{figure}  

Using Eqs. (\ref{2.5}) and (\ref{2.6}) we can now find the kink tail around
both $\phi = 0$ and $\phi = 1$. We note that in order to find the self-dual
kink solution between $0$ and $1$, we need to take $+x$ $(-x)$ in 
Eq. (\ref{3.16}) depending on whether $n$ is an odd (or even) integer. Note 
that this is consistent with what we have used for $n = 1, 2$. 
Using Eq. (\ref{2.6}) we then find that for any integer $n$ 
\be\label{3.17}
\lim_{x \rightarrow -\infty} \phi^{m}(x) \big(\ln[\phi(x)]\big)^{n+1} 
 = (-1)^{n} \frac{1}{mx}\,.
\ee

On the other hand, for both odd and even integer $n$ using Eq. (\ref{2.5})
we find that
\be\label{3.19}
\lim_{x \rightarrow +\infty} \phi(x) = 1
- \frac{1}{\left[nx+\frac{m^{n}\gamma}{(n-1)!}\right]^{1/n}}\,.
\ee 
It is worth pointing out that the asymptotic behaviour around $\phi = 0$
(as $x \rightarrow -\infty$) in Eq. (\ref{3.17}) (i.e. for even integer $n$) 
can also be written as 
\be\label{3.20}
\lim_{x \rightarrow -\infty} \phi(x)^{{\phi(x)}^{m/(n+1)}} 
= e^{1/(mx)^{1/(n+1)}}\,.
\ee
On the other hand the asymptotic behaviour around $\phi = 0$
(as $x \rightarrow -\infty$) in Eq. (\ref{3.19}) (i.e. for odd integer $n$) 
can also be written as 
\be\label{3.21}
\lim_{x \rightarrow -\infty} \phi(x)^{{\phi(x)}^{m/(n+1)}} 
= e^{1/(-mx)^{1/(n+1)}}\,,
\ee
which is known in the literature as the power-tower function of order two \cite{powert} 
or tetration \cite{tetration}. 

\vspace{0.1in}  
\noindent{\bf Kink Mass}

One can easily calculate the kink mass for the entire family of potentials. In particular,
for the kink potential given by Eq. (\ref{3.12}), the kink mass is given by
\be\label{3.16a}
M_K = \int_{0}^{1} \sqrt{2V(\phi)}\, d\phi = \int_{0}^{1} d\phi\, 
\phi^{m+1} [\ln(\phi)]^{n+1} d\phi = \frac{(n+1)!}{(m+2)^2}\,.
\ee
Note that the kink mass decreases as $m$ increases keeping $n$ fixed. On the
other hand the kink mass increases as $n$ increases keeping $m$ fixed. 

\subsection{Stability Analysis}

We can perform the stability analysis of the kink solutions discussed 
in this section and
show that like the previous section (as well as the kinks with the power 
law tail), 
for all the kink 
solutions of this section, there is no gap between the zero mode and the 
onset of the continuum.

As an illustration, we discuss the $m = n = 1$ case in detail, the 
generalization to the arbitrary $m, n$ case is then straightforward.
In the case of $m = n = 1$, the self-dual equation is as given by (\ref{3.3}). 
Thus the kink zero mode is given by
\be\label{3.24}
\psi_0(x) = \frac{d\phi_k}{dx} \propto  [\phi_k(x)]^2 
\big(\ln[\phi_k(x)]\big)^2\,,
\ee
(where $\phi_k$ is the kink solution) which clearly is nodeless and vanishes
as $x \rightarrow \pm \infty$ since as $x$ goes from $-\infty$ to $\infty$,
$\phi$ varies from $0$ to $1$. 

We can also calculate the corresponding potential $V_K(x)$ which occurs in the
stability equation (\ref{2.24}). 
On using the potential for $m = n = 1$ as given by Eq. (\ref{3.2}) we find
that
\be\label{3.25}
V_K(x) = \frac{d^2V(\phi_k)}{d\phi^2} = 6(\phi_k)^2 [\ln(\phi_k)]^4 
+ 14 (\phi_k)^2 [\ln(\phi_k)]^3 + 6 (\phi_k)^2 [\ln(\phi_k)]^2\,, 
\ee
and hence it is clear that $V_K(x = \infty) = V_K(-\infty) = 0$.
Thus the continuum begins at $\omega^2 = 0$, i.e. there is no gap between
the zero mode and the onset of the continuum.

The generalization to the arbitrary $m, n$ case is now straightforward. In
particular, we find that
the kink zero mode in that case is given by
\be\label{3.26}
\psi_0(x) = \frac{d\phi_k}{dx} = [\phi_k(x)]^{m+1}(x) 
\big(\ln[\phi_k(x)]\big)^{n+1}\,,
\ee
which is clearly nodeless. The corresponding kink stability potential
is given by
\begin{eqnarray}\label{3.27}
V_K(x) &=& \frac{d^2V(\phi_k)}{d\phi^2} = (m+1)(2m+1)[\phi_k(x)]^{2m} 
[\ln(\phi_k)]^{2n+2} \\ \nonumber
&+& (4m+3)(n+1) [\phi_k(x)]^{2m} [\ln(\phi_k)]^{2n+1} 
+(n+1)(2n+1) (\phi_k)^{2m} [\ln(\phi_k)]^{2n}\,, 
\end{eqnarray}
from where again it is clear that there is no gap between the zero mode and the
onset of the continuum. 

\subsection{Nature of Kink-Kink and Kink-Antikink Interactions} 

In this section we have obtained the kink  and the mirror kink solutions (and 
the corresponding antikinks) for  the two-parameter family
of potentials as given by Eq. (\ref{3.1}). The kinks are from $0$ to $1$ (and
the mirror kinks are from $-1$ to $0$) as $x$ goes from $-\infty$ to $\infty$.
We have seen that while around $\phi = 1$ or $\phi = -1$ one has a power law 
tail, the kink tail around $\phi = 0$ has a power-tower form. 
Using this information we attempt to qualitatively understand the nature of
the KK and the K-AK interactions. 

Let us first consider the kink-kink interaction between the $(-1,0)$ mirror 
kink and the $(0,1)$ kink. From Eq. (\ref{2.1}) it is clear that around 
$\phi = 0$ the kink potential is as given by Eq. (\ref{3.1}). Notice that
if there were no $[\ln(\phi^2)]^{2n+2}$ term in Eq. (\ref{3.1}) then using the 
recent
approach of Manton for potentials with a power law tail \cite{man, ivan} one
would have immediately predicted that the KK force would be as given by 
Eq. (\ref{2.25}). 
So the question is: what is 
the effect of the $[(1/2)\ln(\phi^2)]^{2(n+1)}$ term multiplying the 
potential in 
Eq. (\ref{3.1})? In this connection as noted earlier,  
in a recent publication 
\cite{KKS} we have shown that in the case of the potential
$V(\phi) = (1/2) \phi^2 [(1/2)\ln(\phi^2)]^2$, while the kink-kink
force would have been exponentially small if there were no $\ln(\phi^2)$ term 
present, because of the $\ln(\phi^2)$ term, the force actually gets even weaker
and is in fact super-exponential. Taking this as a guide, we would 
therefore expect that the KK force in the case of the potential (\ref{3.1}) 
will still have a power law fall off but perhaps with a slower fall off and the 
strength of 
the force too would be different. Only either a new 
formalism or numerical estimation can decide the issue. 

Similar conclusion is also true concerning the force between $(1,0)$ AK and
$(0,1)$ K. On the other hand the force between $(0,1)$ K and $(1,0)$ 
AK can be immediately estimated using the recent Manton formalism \cite{man, ivan} 
and is given by
\be\label{3.28}
F_{KK} = \bigg [-\frac{\sqrt{\pi} \Gamma[n/2(n+1)]}
{\Gamma[-1/(2(n+1)]} \bigg ]^{2(n+1)/n} \frac{1}{2 R^{2(n+1)/n}}\, 
\ee 
where $R$ denotes the distance between a kink and an antikink.  

\section{Summary}

In this paper we have considered a continuous one-parameter family of 
potentials as given by Eq. (\ref{2.1}), all of which have kink tails
of the form $ette$ where $e$ and $t$ correspond to exponential and power-tower 
type of tail, respectively. Similarly, in Sec. III we have constructed a 
two-parameter family of potentials given by Eq. (\ref{3.1}), all of which admit 
kink tails of the form
$pttp$ where $p$ corresponds to power law type of tail. For all these cases we have 
calculated the corresponding kink masses. Further, we have shown that the kink
stability equation in all these cases is such that there is no gap between the
zero mode and the beginning of the continuum in the Schr\"odinger-like 
equation. Finally, 
we have also qualitatively discussed the nature of the kink-kink and 
the kink-antikink interactions in all these cases. 

By now we have a large number of kink bearing models which admit
kinks with a variety of tails such as exponential \cite{Raj}, power-law 
\cite{KCS, Chapter, gomes, bazeia}, super-exponential \cite{KKS, menezes, 
marques, bazeia2} and power-tower. It is then natural to enquire if there is a 
recipe for constructing models which admit such a diverse variety of kink tails.
In this context we might add that the recipe for constructing kink solutions
with an exponential or a power law tail is well known \cite{KS}. For 
completeness we mention it first and then give the recipe for constructing 
the kink solutions with either super-exponential or power-tower type of tail. 

Since a kink has finite energy it implies that the solution must approach 
one of the minima (vacua), say $\phi_0$, of the theory as 
$x \rightarrow \pm \infty$. If the lowest non-vanishing derivative of the
potential at the minimum has order $m$, then by Taylor series expansion of the potential
at the minimum and writing the field close to it as $\phi = \phi_0 + \eta$,
one finds that the self-dual first order equation in $\eta$ implies that
(assuming that the potential vanishes at the minimum) 
\be\label{4.1}
\frac{d\eta}{dx} \propto \eta^{m/2}\,.
\ee
Thus if $m = 2$ then $\eta \propto e^{-\alpha x}$ (i.e. exponential tail) while if 
$m > 2$ then $\eta \propto 1/x^{2/(m-2)}$ (i.e. power law tail). 

In our recent paper about the super-exponential
tail \cite{KKS}, we have shown that if instead 
\be\label{4.2}
\frac{d\eta}{dx} \propto \eta^2 \ln(\eta^2)\,,
\ee
then $\eta \propto e^{-e^{-\alpha x}}$, so that there is a super-exponential tail.

On the other hand using the results of this paper it is clear that if 
\be\label{4.3}
\frac{d\eta}{dx} \propto \eta^{m+1} [\ln(\eta^2)]^{n+1}\,, ~~m, n \ge 1\,,
\ee
then $\eta$ is a solution of the equation
\be\label{4.4}
\eta^{m} [\ln\eta]^{n+1} = (-1)^{n} \frac{1}{mx}\,, 
\ee  
which leads to power-tower kink tails. 

\begin{figure}[h] 
\includegraphics[width= 5.1 in]{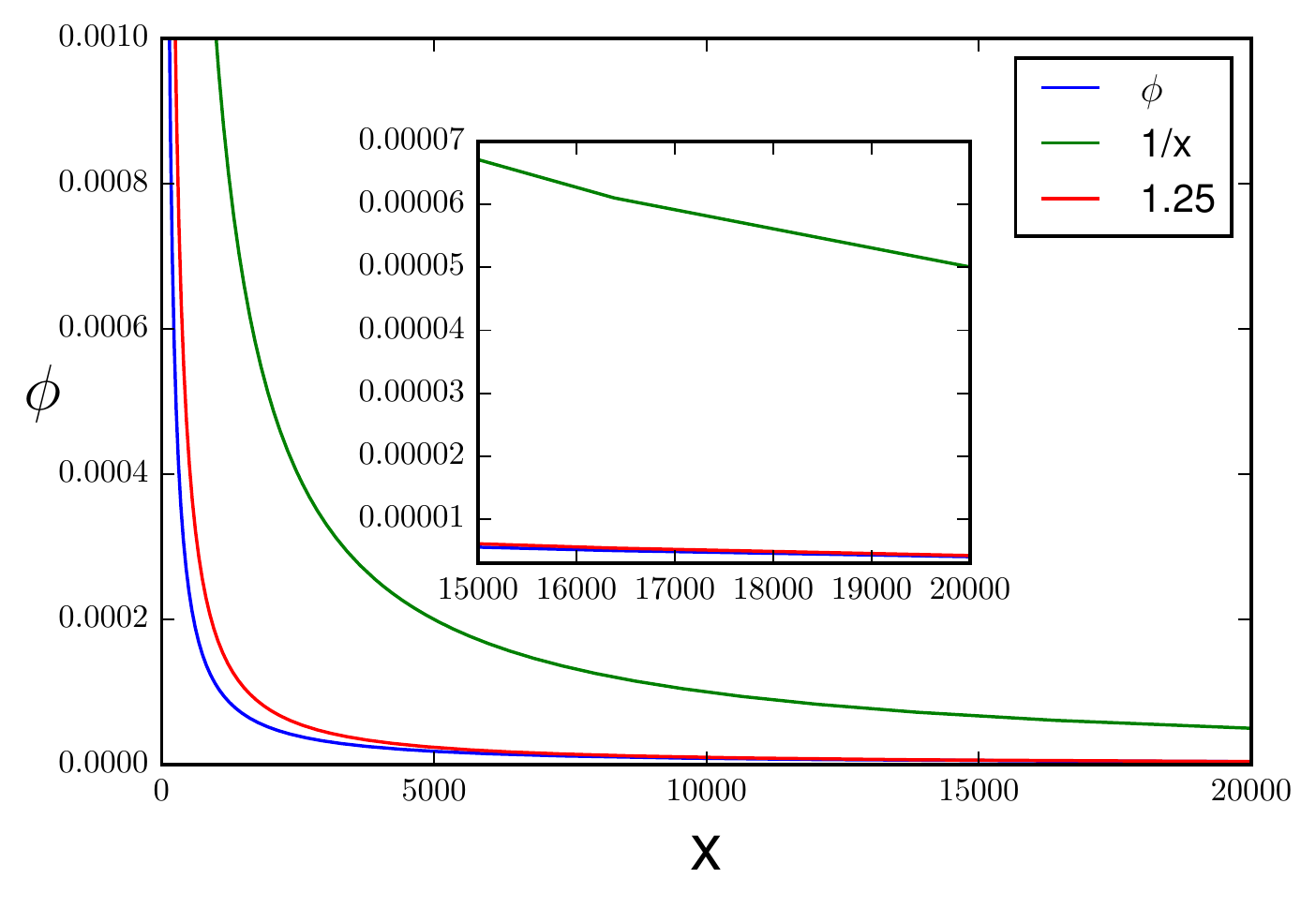}
\caption{$\phi(x)$ vs $x$ obtained by inverting Eq. (58) and compared with
$\phi = 1/x$ and $\phi = 1/x^{1.25}$. }
\end{figure} 

\begin{figure}[h] 
\includegraphics[width= 5.1 in]{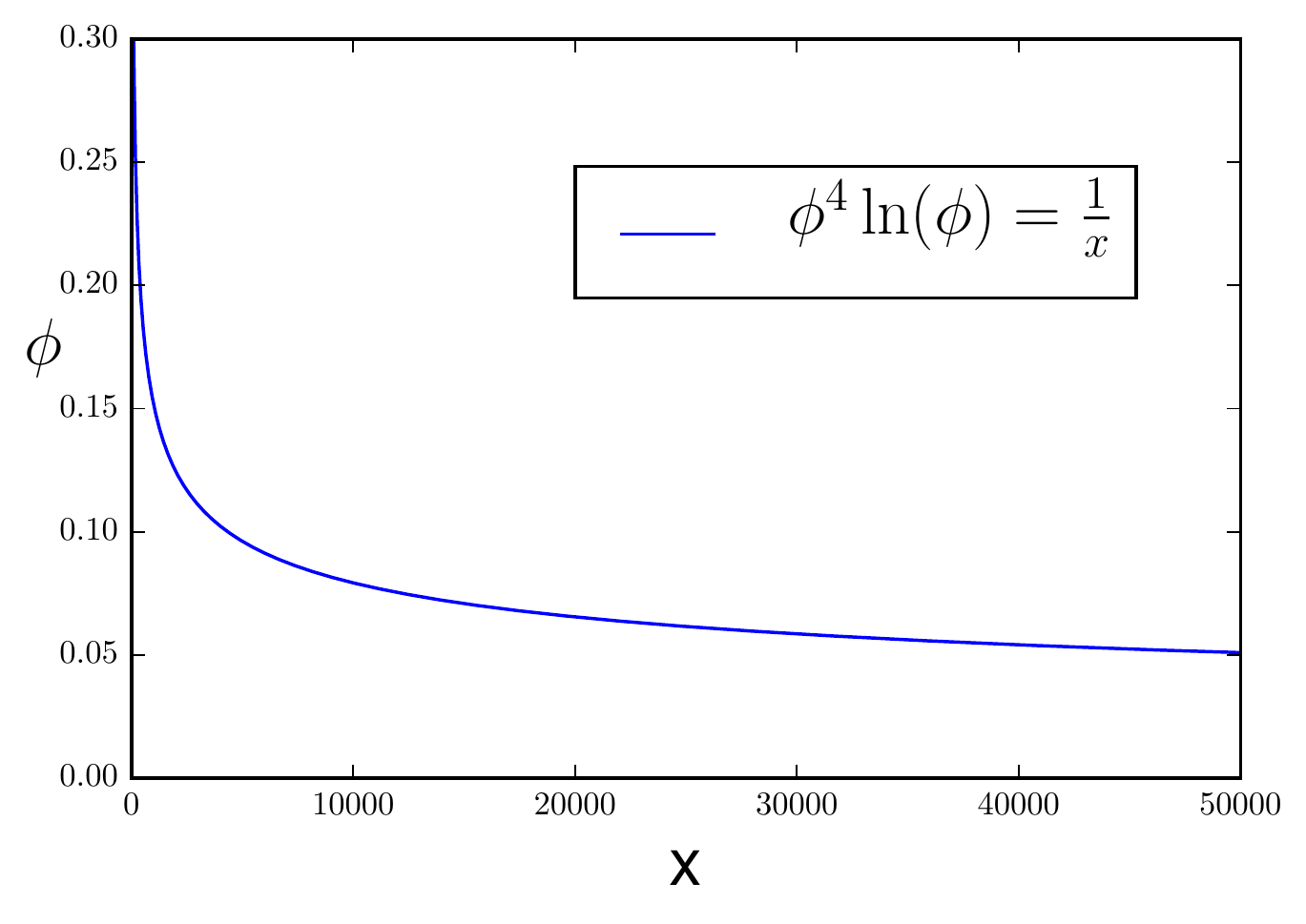}
\caption{$\phi(x)$ vs $4x$ obtained by inverting Eq. (60) for the case $m=4$. }
\end{figure}

Before ending this discussion, it is worth pointing out some of the open 
problems in the context of power-tower type kink tails. 

\begin{enumerate}

\item In this paper we have constructed one- and two-parameter family of 
potentials which lead to kink 
tails of the form $ette$ and $pttp$, respectively. The obvious question is, 
can one similarly construct at least a one-parameter family of potentials 
which gives tails of the form
$teet$, $tppt$, $tttt$ as well as the mixed tails of the form $ettp$, $eppt$ 
and $peet$? Finally, can one construct models with an admixture of 
super-exponential tails and exponential and/or power law and/or power-tower 
type of tails? 

\item In this paper we have not been able to explicitly calculate the 
force between the $(-1, 0)$ kink and the $(0,1)$ kink
since the two ends facing each other have power-tower type of tails and
in this case it is not straightforward to invert and obtain the behaviour 
of the 
tail as a function of $x$ when $x \rightarrow -\infty$. As an illustration,
consider the asymptotic behaviour around $\phi = 0$ in case $x \rightarrow
-\infty$ as given by Eq. (\ref{2.7}), that is 
\be\label{4.5}
\lim_{x \rightarrow -\infty} \phi(x) \ln(\phi(x)) = \frac{1}{x}\,.
\ee 
If $\ln(\phi(x))$ were not there then we know that for large negative $x$, 
$\phi(x) \propto -1/x$.  
In this connection we notice that in a recent publication 
\cite{KKS} we have shown that in the case of the potential
$V(\phi) = (1/2) \phi^2 [(1/2)\ln(\phi^2)]^2$, while the kink tail around 
$\phi = 0$ would have been an exponential tail 
in case  there were no $\ln(\phi^2)$ term 
present, because of the $\ln(\phi^2)$ term, the kink tail actually gets even 
weaker and is in fact super-exponential. Taking this as a guide, we
speculate that corresponding to the power-tower form as given by 
Eq. (\ref{4.5}),
the behaviour of $\phi(x)$ for large negative $x$ should be of the form
\be\label{4.6}
\lim_{x \rightarrow -\infty} \phi(x) = 
\frac{1}{(-x)^{1+\epsilon_{1, 0}}}\,,
~~\epsilon_{1,0} > 0\,.
\ee 
Here by $\epsilon_{1, 0}$ one means $\epsilon_{m = 1, n = 0}$ 
corresponding to $\phi^{m}$ and $[\ln(\phi)]^{n+1}$ in Eq. (\ref{4.5}) with
$m = 1, n = 0$. We have 
inverted Eq. (\ref{4.5}) numerically and from Fig. 7 we see that 
it can be
fitted in the form of Eq. (\ref{4.6}) with $\epsilon_{1, 0}$ 
approximately equal to $0.25$. This would imply that the potential around 
$\phi = 0$ is of the form $\phi^{2k}$ with $k = 9/5$. We might add here that
the new Manton formalism \cite{man, ivan}, even though developed for integral 
$k$ is also valid for any real number $k$. Using this information, one can 
estimate the
force between the $(-1,0)$ K and the $(0,1)$ K using the new Manton formalism 
and show
that the KK force would vary like $R^{-9/2}$, where $R$ is the distance between
the two kinks.

In the same way, one can numerically invert for any $m$ the equation
around $\phi = 0$ for large negative $x$ as given by Eq. (\ref{2.16}), i.e.  
\be\label{4.7}
\lim_{x \rightarrow -\infty} \phi(x)^{m} \ln(\phi(x)) = \frac{1}{mx}\,,
\ee 
and try to numerically estimate the corresponding exponent. As an illustration, 
in Fig. 8 we have inverted Eq. (\ref{4.7}) for the case of $m = 4$.

\item Generalization of the above discussion in the case of the power-tower 
Eq. (\ref{3.17}) with arbitrary $m$ and $n$ 
is now straightforward. In particular, if the power-tower equation is given by
\be\label{4.8}
\lim_{x \rightarrow -\infty} \phi^{m}(x) [\ln(\phi(x)]^{n+1} 
= (-1)^{n} \frac{1}{mx}\,,
\ee 
then we speculate that the behaviour of $\phi(x)$ for large negative $x$ 
should be of the form
\be\label{4.9}
\lim_{x \rightarrow -\infty} \phi(x) = 
\frac{1}{(-mx)^{1/m +\epsilon_{m,n}}}\,,
~~\epsilon_{m,n} > 0\,.
\ee
As an illustration, in Fig. 9 we have numerically inverted Eq. (\ref{4.8}) 
in case
$m = 1, n = 1$ and we find that $\epsilon_{1, 1}$ is approximately equal 
to 0.5 which is larger than $\epsilon_{1, 0}$, which is approximately 0.25.

In case the exponent is 0.5, this would imply that the potential around 
$\phi = 0$ is of the form $\phi^{2k}$ with $k = 5/3$. 
Using this information, one can estimate the
force between $(-1,0)$ K and $(0,1)$ K using the new Manton formalism 
\cite{man, ivan} and show
that the KK force would go like $R^{-5}$, where $R$ is the distance between
the two kinks.

\begin{figure}[h] 
\includegraphics[width= 5.1 in]{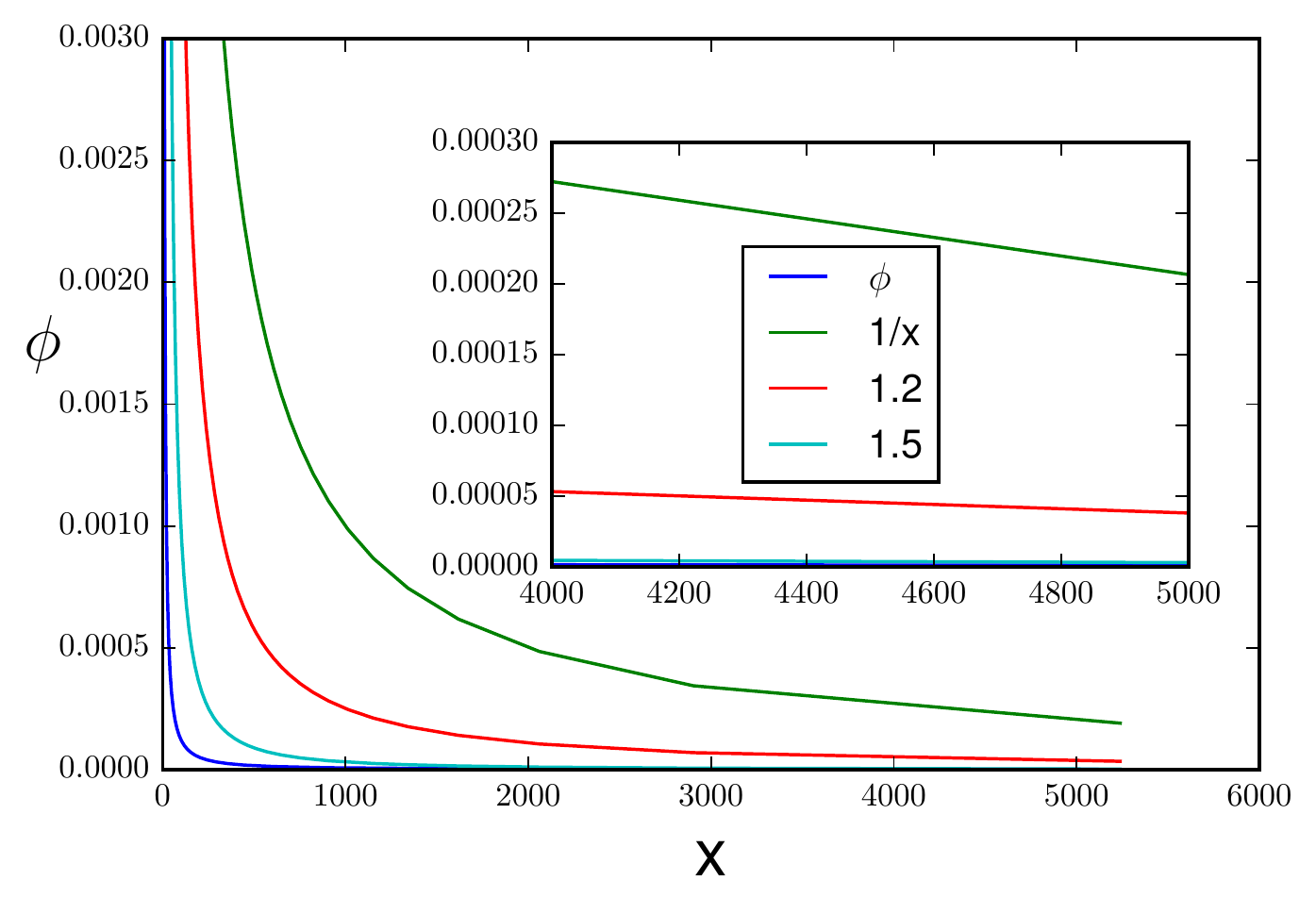}
\caption{$\phi(x)$ vs $x$ obtained by inverting Eq. (61) for the case $m=1$ and $n=2$,  
and compared with $\phi=1/x$,  $\phi=1/x^{1.2}$ and $\phi=1/x^{1.5}$. }
\end{figure}  

\begin{figure}[h] 
\includegraphics[width= 5.1 in]{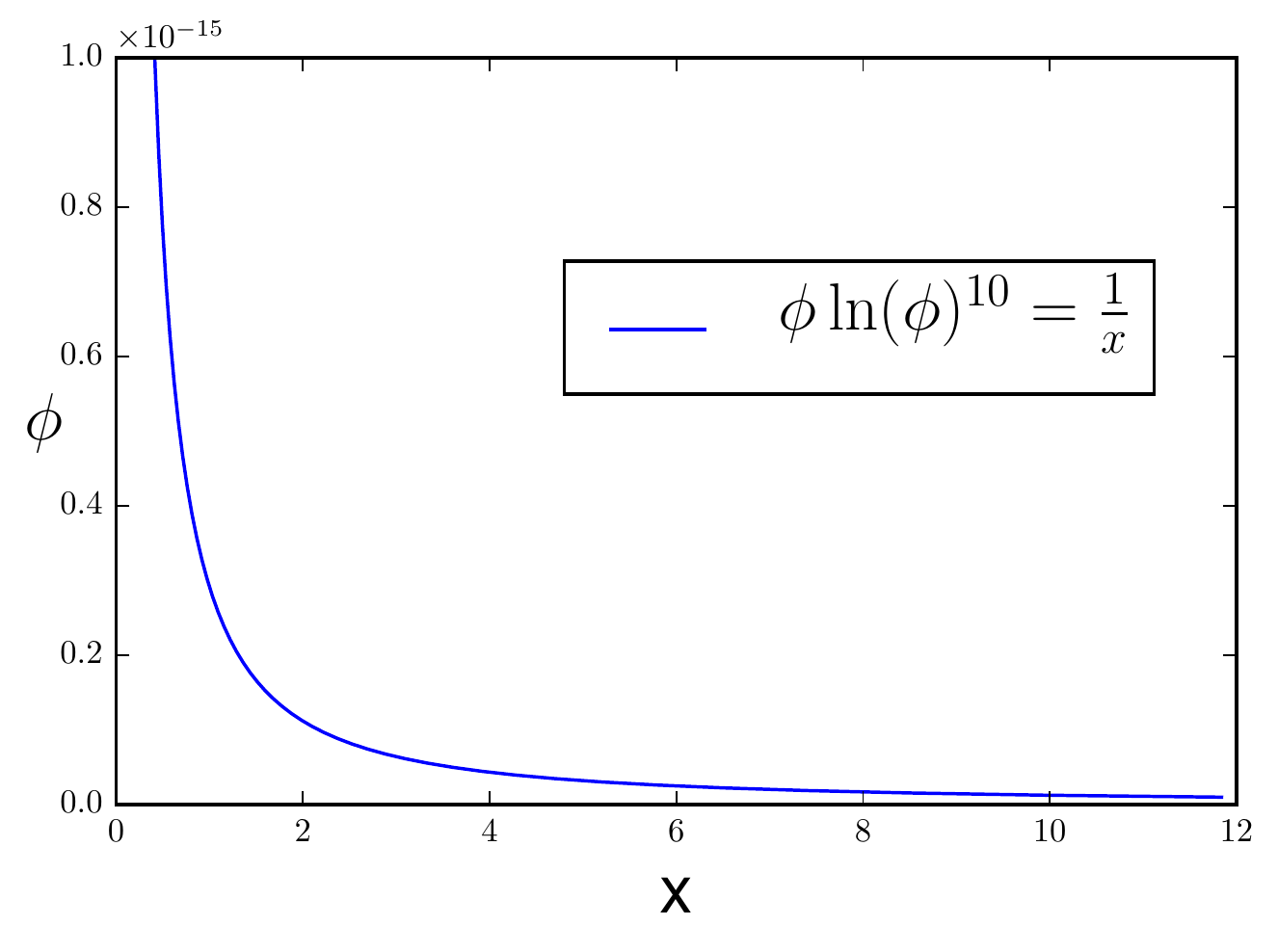}
\caption{$\phi(x)$ vs x obtained by inverting Eq. (61) with $m = 1$ and $n = 9$. }
\end{figure}   

\item Looking at the two examples of $m = 1, n = 0$ and $m = 1, n = 1$ it
is immediately clear that $\epsilon_{1, 0} < \epsilon_{1, 1}$. We speculate 
that in general for arbitrary $m, n$ we will have the inequality
$\epsilon_{m,n_1} < \epsilon_{m, n_2}$ in case $n_1 < n_2$.

\item In the same way, one can invert for any $n$ and $m$ the equation
around $\phi = 0$ for large negative $x$ as given by Eq. (\ref{4.8}). 
As an illustration, in Fig. 10 we have numerically inverted Eq. (\ref{4.8}) 
in case $m = 1, n = 9$. It is clear from the figure that the exponent
$\epsilon_{1, 9}$ is significantly bigger than 0.5. It thus appears that as 
$n$ becomes progressively larger, effectively the kink tail around $\phi = 0$ 
for large negative $x$ will approach an exponential tail.

We first elaborate our argument in the case of $m = 1$ and arbitrary $n$. 
Generalization to arbitrary $m$ and $n$ is then straightforward. For $m = 1$ 
and arbitrarty $n$ case, the corresponding exponent is $\epsilon_{1,n}$. 
This would imply that the potential around 
$\phi = 0$ is of the form $\phi^{2k}$ with 
\be\label{4.10} 
k = \frac{2+\epsilon_{1,n}}{1+\epsilon_{1,n}}\,. 
\ee
Now we have seen from the examples of $m = 1, n = 0$; $m = 1, n = 1$ and 
$m = 1, n = 10$ that as $n$ increases $\epsilon_{1, n}$ becomes progressively 
larger. 
In other words, for very large 
$n$ we expect that $\epsilon_{1, n} \gg 2$ and hence  
for very large $n$, $k$ as defined by Eq. (\ref{4.10}) tends to $1$ which
corresponds to an exponential tail. However, we would like to emphasize that
no matter how large $n$ is, so long as it is finite, $k$ is strictly greater
than one such that for all finite $n$, the kink tail has power law fall off 
thereby justifying the name power-tower. 
Therefore, kinks with power-tower type of tails provide a bridge between kinks 
with power law type of tails and kinks with exponential tails. 

Generalization to arbitrary $m$ is now straightforward. In this case the
corresponding exponent is $\epsilon_{m, n}$. This would imply that the
potential around $\phi = 0$ is of the form $\phi^{2k}$ with 
\be\label{4.11} 
k = \frac{m+1 +m \epsilon_{m,n}}{1+m \epsilon_{m,n}}\,. 
\ee
We surmise that for very large $n$, no matter what $m$ is, $m \epsilon_{m, n} \gg m+1$, 
so that even in this case $k$ would tend to one which corresponds to an exponential tail, 
although for any large but finite $m, n$, it will strictly be greater than one. 

Also as shown in Sec. III, for arbitrary $n$ and $m$, the
kink stability equation is such that there is no gap between the zero mode
and the beginning of the continuum, which is the hallmark of kink solutions
with a power law tail. 

It is worth pointing out that since for arbitrary $m$ and $n$, the potential 
around $\phi = 0$  is of the form $\phi^{2k}$ with $k$ as given by 
Eq. (\ref{4.11}) hence using the new Manton formalism \cite{man, ivan} 
one can show that in that case the KK force would go like $R^{-d}$,  where    
$d = 2 [1+\epsilon_{m,n} +1/m]$.

\item In this paper by numerically inverting power-tower equations in a few 
cases we have shown that the exponent $\epsilon_{1,n}$ satisfies the inequality
$\epsilon_{1, 0} < \epsilon_{1, 1}$. From here it
is natural to speculate that even for arbitrary $m$, the exponents will
satisfy the inequality $\epsilon_{m, n_1} < \epsilon_{m, n_2}$ in case 
$n_1 < n_2$. It would be desirable if one can prove this rigorously. 

One of the obvious open problems is about the exponent for 
fixed $n$ but varying $m$. In particular, what can one say about
$\epsilon_{m_1, n} - \epsilon_{m_2, n}$ in case $m_1 < m_2$? Similarly 
proceeding further, can one say anything about $\epsilon_{m_1, n_2} - 
\epsilon_{m_2, n_1}$ where $n_1 < n_2$ and $m_1 < m_2$?

\end{enumerate}

We hope to address some of these issues in the near future. 

\vskip 0.1 in
\noindent{\bf Acknowledgement:}  We acknowledge useful discussions with Ayhan Duzgun. A.K. is grateful to Indian National Science Academy (INSA) for 
the award of INSA Scientist position at Savitribai Phule Pune University. This work was supported in part by the U.S. Department of Energy.

\end{document}